\documentclass[10pt,letterpaper]{article}
\usepackage{opex3}
\usepackage{color}
\usepackage{amsmath}
\usepackage{amssymb}
\usepackage{upgreek}
\usepackage{cite}

\begin{document}

\title{A tunable CW UV laser with $<$35\,kHz absolute frequency instability for precision spectroscopy of Sr Rydberg states}

\author{Elizabeth~M.~Bridge$^*$, Niamh~C.~Keegan, Alistair~D.~Bounds, Danielle~Boddy, Daniel~P.~Sadler, and Matthew~P.~A.~Jones}

\address{Joint Quantum Centre (JQC) Durham-Newcastle, Department of Physics, Durham University, Durham DH1 3LE, UK}

\email{$^*$elizabeth.bridge@durham.ac.uk} 

\begin{abstract}
We present a solid-state laser system that generates over 200\,mW of continuous-wave, narrowband light, tunable from 316.3\,nm~--~317.7\,nm and 318.0\,nm~--~319.3\,nm. The laser is based on commercially available fiber amplifiers and optical frequency doubling technology, along with sum frequency generation in a periodically poled stoichiometric lithium tantalate crystal. The laser frequency is stabilized to an atomic-referenced high finesse optical transfer cavity. Using a GPS-referenced optical frequency comb we measure a long term frequency instability of $<$\,35\,kHz for timescales between $10^{-3}$\,s and $10^3$\,s. As an application we perform spectroscopy of Sr Rydberg states from $n\,=\,37$~--~81, demonstrating mode-hop-free scans of 24\,GHz. In a cold atomic sample we measure Doppler-limited linewidths of 350\,kHz. 
\end{abstract}



\section{Introduction}
Tunable continuous-wave (CW) lasers are the workhorse for quantum technologies based on ultracold atoms and ions. While the visible and near-infrared spectral regions are readily accessible by solid-state laser systems, the extension of these techniques to the near-ultraviolet region remains an active area of research. Many ionic and atomic species have their principal cooling transitions in the near-UV, including species such as Be$^+$ \cite{Bollinger1985} and Hg \cite{Hachisu2008} that are relevant for optical frequency standards.

In addition to laser cooling, narrowband CW excitation enables precision spectroscopy of highly excited Rydberg states. Rydberg atoms have high principal quantum number $n$ and exhibit large atomic polarizabilities, which scale as $n^{7}$, thus providing tunable long-range interactions and high sensitivity to electric fields. With sufficiently narrowband excitation the long-range interactions between pairs of Rydberg atoms can be observed. In particular, if the interaction shift exceeds the linewidth of the excitation, then a blockade effect occurs \cite{Lukin2001} and a single excitation is shared between all the atoms within a blockade radius $R_{\mathrm{B}} = \left(C_6/\hbar \Omega \right)^{1/6}$, where $C_6$ is the van der Waals interaction strength and $\Omega$ is the excitation Rabi frequency. 

The Rydberg blockade effect has been used to demonstrate quantum effects in neutral atoms, including cooperativity \cite{Pritchard2010}, entanglement \cite{Wilk2010}, a controlled-NOT gate \cite{Isenhower2010} and coherent many-body Rabi oscillations \cite{Dudin2012}. Other applications of precision Rydberg spectroscopy include electrometry \cite{Abel2011, Sedlacek2012, Lochead2013}, production of single photons \cite{Saffman2002} and the creation of ultra-long range molecules \cite{Greene2000, Boisseau2002, Bendkowsky2009,DeSalvo2015}. By optically ``dressing'' the ground or a metastable state with a Rydberg state, further potential applications include demonstration of a super-solid phase of matter \cite{Henkel2010,Cinti2010} and spin-squeezing in optical lattice clocks for increased signal-to-noise beyond the standard quantum limit \cite{Gil2014}. Keating \textit{et al.} \cite{Keating2015} have recently demonstrated Rydberg dressing in Cs atoms for studying low-decoherence implementation of a controlled-Z gate. 

So far, most of these experiments have relied on a two-step excitation scheme via a short-lived, low-lying intermediate state. While this has the advantage that both lasers may be in the visible or near-IR spectral regions, the short lifetime of the intermediate state introduces additional decoherence. Electromagnetically induced transparency \cite{Mohapatra2007} or off-resonant two-photon excitation \cite{Deiglmayr2006} can reduce this effect, but at the expense of a lower Rydberg excitation efficiency. More recently, experiments in the alkali atoms have used direct excitation from the ground state, which requires tunable CW light at 297\,nm (Rb) \cite{Thoumany2009} and 319\,nm (Cs) \cite{Hankin2014}. 

An alternative approach to reducing the decoherence of the two-photon excitation scheme is provided by two-electron atoms such as Sr. Here, intercombination transitions give access to long-lived intermediate states. These provide a coherent two-photon excitation route to the triplet Rydberg states, including the 5s$n$s\,$^3$S$_1$ series which have near-isotropic wavefunctions and interaction strengths \cite{Vaillant2012}. Isotropic interactions relax the geometric constraints of the atomic sample for many of the proposed Rydberg and dressing experiments, and could be key for demonstrating a supersolid phase of matter \cite{Henkel2010,Cinti2010}. Moving to two-electron atoms also enables applications, such as spin-squeezing \cite{Gil2014} and blackbody thermometry \cite{Ovsyanikov2011}, in optical lattice clocks. In Sr this requires tunable radiation in the 316\,nm~--~322\,nm region [Fig.~\ref{fig:fig1}(a)]. Recently, the first experiments using this intercombination excitation route in Sr have led to the observation of Sr Rydberg molecules \cite{DeSalvo2015} and the study of Autler-Townes spectra in a dense Sr Rydberg gas \cite{DeSalvo2015b}. Combined with the ability to optically manipulate the second valence electron for trapping \cite{Mukherjee2011}, state detection \cite{Millen2010,Millen2011} and imaging \cite{Lochead2013}, intercombination line Rydberg excitation provides a powerful toolbox for exploring many of these ideas. 

The conventional approach to generating tunable CW light in the near-UV region of the spectrum has been to frequency-double a dye laser \cite{Thoumany2009,Beigang1982b,Manthey2014}. Solid-state systems using sum-frequency generation (SFG) have been developed for particular cooling transitions, offering high power and narrow linewidth \cite{Hankin2014,Wilson2011,Rengelink2015}, but a tuning range of only a few GHz. Recently these techniques have been extended to lasers for Rydberg spectroscopy \cite{Hankin2014}, where the restricted tuning range means that only a narrow range of principal quantum numbers are accessible.

In this paper we describe a solid-state, narrow-linewidth UV laser system tunable from 316.3\,nm~--~317.7\,nm and 318.0\,nm~--~319.3\,nm, which enables the two-step excitation of Sr Rydberg states with principal quantum number in the range $n=35\rightarrow$\,$>$\,300 from two different intermediate states that are separated by 5.6\,THz. We achieve a stable output power of $>200\,$mW. A dual stabilization scheme based on a high-finesse tunable optical transfer cavity and an atomic reference provides a long-term absolute frequency instability of $<35\,$kHz, measured relative to a GPS-stabilized optical frequency comb. The utility of the laser for Rydberg physics is demonstrated by the spectroscopy of high-lying triplet Rydberg states in Sr. 

\section{\label{sec:laser}Laser construction and characterization}
\begin{figure}
	\includegraphics[width=\columnwidth]{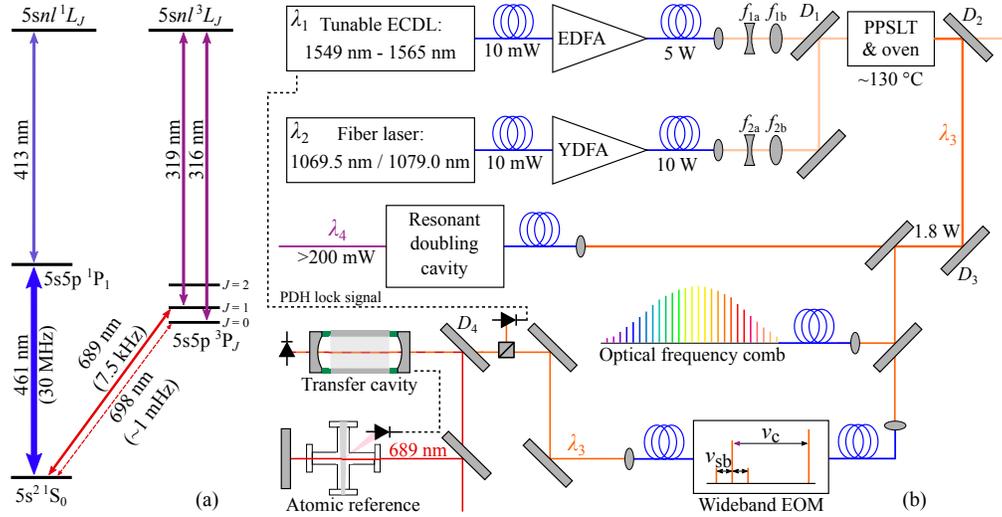}
	\caption{(a) Energy level diagram of relevant transitions in atomic Sr. The primary cooling transition at 461\,nm is used for Zeeman slowing and cooling the atoms in a magneto-optical trap (MOT). Two-photon excitation with the 461\,nm and 413\,nm lasers drives atoms up to the singlet Rydberg series, as was used for our previous work \cite{Lochead2013,Millen2010,Millen2011}. The second-stage cooling transition at 689\,nm is used to cool the atoms to $\sim\,1$~--~$10\,\upmu$K in a MOT. Two-photon excitation with 689\,nm and 319\,nm, or 698\,nm and 316\,nm drives the atoms up to the triplet Rydberg series. (b) Schematic of the laser system. A PPSLT crystal is used to sum the frequencies of two infra-red lasers at wavelengths of $\lambda_1$ and $\lambda_2$. The resulting light at $\lambda_3$ is frequency doubled to produce $>\,200\,$mW in the UV ($\lambda_4$). The laser frequency is locked to an optical transfer cavity stabilized to the 5s$^2$\,$^1$S$_0$~$\rightarrow$~5s5p\,$^3$P$_1$ intercombination line in Sr at 689\,nm. A wideband electro-optic modulator (EOM) is used to bridge the frequency gap between the cavity mode and the Rydberg transition. The laser frequency is measured on a GPS-referenced optical frequency comb. (EDFA = Er-doped fiber amplifier, YDFA = Yb-doped fiber amplifier. $D_1 - D_4$ = dichroic mirrors). }
\label{fig:fig1}
\end{figure}

The goal was to develop a laser system suitable for exciting to a large range of Rydberg states in the triplet series of Sr, via two different intermediate states [Fig.~\ref{fig:fig1}(a)]. A narrow laser linewidth is required in order to achieve a large Rydberg blockade radius, and high power is necessary for off-resonant Rydberg dressing since the coupling matrix elements are weak. With these criteria in mind, we have built a high power, narrow-linewidth, widely-tunable laser system for excitation to the triplet Rydberg series from both the 5s5p\,$^3$P$_0$ and 5s5p\,$^3$P$_1$ intermediate states. The full range of wavelengths achievable, and the corresponding Rydberg states that can be addressed, are given in Table~\ref{tab:crystal}. 
 
Our approach, shown in Fig.~\ref{fig:fig1}(b), is based on that described in \cite{Wilson2011}, and is similar to that used in \cite{Hankin2014,Rengelink2015}. Two infra-red seed lasers are amplified and combined using sum frequency generation (SFG) in a periodically-poled stoichiometric lithium tantalate (PPSLT) crystal to produce red light at $\sim$\,633\,nm~--~639\,nm. Subsequently, UV light is produced using resonantly-enhanced second harmonic generation (SHG). A small amount of the red light is separated off for locking to an atomic-referenced optical cavity, and for independent absolute frequency measurements using a GPS-referenced optical frequency comb.

The key challenge compared to previous work was to extend the SFG technique to meet our tuning requirements. Firstly, we exploit the wide gain bandwidth of Er- and Yb-doped fiber amplifiers to achieve high fundamental power over a wide range of wavelengths. We employ two commercial fiber amplifier systems; an Er-doped amplifier (Manlight) that produces up to 5\,W over the wavelength range $\lambda_1$\,=\,1549\,nm to 1565\,nm, and a Yb-doped amplifier (Nufern) that produces up to 10\,W over $\lambda_2$\,=\,1064\,nm to 1083\,nm. Table \ref{tab:crystal} summarizes how we exploit these wavelength ranges to address the Rydberg series of interest. 
\begin{table}[h]
\caption{Wavelengths achievable with our laser system and the corresponding principal quantum numbers $n$ of the Sr Rydberg states we can access. Wavelengths $\lambda_{1}$ and $\lambda_{2}$ are combined using SFG to produce $\lambda_3$, which is frequency doubled to give the desired wavelength, $\lambda_4$. }
\centering
\begin{tabular}{cccccc}
\hline
\begin{tabular}[c]{@{}c@{}}$\lambda_1$ \\ (nm) \end{tabular} &
\begin{tabular}[c]{@{}c@{}}$\lambda_2$ \\ (nm) \end{tabular} &
\begin{tabular}[c]{@{}c@{}}$\lambda_3$ \\ (nm) \end{tabular} &
\begin{tabular}[c]{@{}c@{}}$\lambda_4$ \\ (nm) \end{tabular} &
\begin{tabular}[c]{@{}c@{}}Intermediate\\ state \end{tabular} &
\begin{tabular}[c]{@{}c@{}}Rydberg\\ $n$ \end{tabular}  \\ \hline
1549.0~--~1565.0 & 1069.5 & 632.7~--~635.3 & 316.3~--~317.7 & $^3$P$_0$ & 32~--~320  \\
1549.0~--~1565.0 & 1079.0 & 636.0~--~638.7 & 318.0~--~319.3 & $^3$P$_1$ & 35~--~$\infty$ \\
\hline
\end{tabular}
\label{tab:crystal}
\end{table}

The 5.6\,THz frequency shift required to change between the two intermediate states is achieved by switching between two different seed lasers centered at $\lambda_{\mathrm{2a}}\,=$\,1069.5\,nm, or $\lambda_{\mathrm{2b}}\,=$\,1079.0\,nm. Both seed lasers are narrow-linewidth ($\sim$\,10\,kHz), high stability fiber lasers (NP Photonics) with polarization-maintaining single-mode optical fiber outputs. The advantage of switching the seed laser in this way is that it requires no significant optical realignment. Tunability within each Rydberg series is achieved by using a highly tunable external cavity diode laser (ECDL) (Toptica Photonics) as the other seed laser ($\lambda_1$). The laser linewidth is somewhat broader than available fiber laser technology ($\sim$\,100\,kHz), but the frequency can be tuned over the whole usable gain bandwidth of the fiber amplifier. 

The final step is to ensure that the SFG process is quasi-phase-matched across these wavelength ranges. The phase mismatch $\Delta k(T)$ can be calculated \cite{Myers1997} by considering the extraordinary refractive index $n_{e,\lambda_i}(T)$ of the crystal \cite{Bruner2003} for a particular temperature $T$ and vacuum wavelength $\lambda_i$ (where $i=1,2,3$ corresponds to the two input, and one output, beams respectively), and the poling period of the crystal $\Lambda_c(T)$:
\begin{equation}
\frac{\Delta k\left(T\right)}{2\pi} = \frac{n_{e,\lambda_3}\left(T\right)}{\lambda_3} - \frac{n_{e,\lambda_2}\left(T\right)}{\lambda_2} - \frac{n_{e,\lambda_1}\left(T\right)}{\lambda_1} - \frac{1}{\Lambda_c(T)}\mbox{.}
\label{eqn:temp1}
\end{equation}
We considered both PPSLT and periodically-poled lithium niobate (PPLN) crystals. The latter offers a higher nonlinear coefficient ($d_{33\mathrm{LT}}/d_{33\mathrm{LN}} \approx 0.67$) \cite{Hum2007}, at the expense of higher light-induced infra-red absorption and a lower photo-refractive damage threshold \cite{Hum2007,Kitamura2001}, and therefore we chose to work with PPSLT. Inserting the relevant material parameters \cite{Bruner2003,Kim1969} in to Eq.~(\ref{eqn:temp1}) we find that five equally spaced poling periods between 13.05\,$\upmu$m and 13.45\,$\upmu$m will provide quasi-phase-matching across the full Rydberg series for crystal temperatures in the range 100\,$^{\circ}$C~--~140\,$^{\circ}$C.

The PPSLT crystal we use (Laser 2000) has a length of $L$\,=\,50\,mm and a thickness of 1\,mm. Distributed across its 15\,mm width are the five independent periodically-poled channels. The crystal is mounted in a temperature stabilized oven (Laser 2000), which allows for fine tuning of the refractive index and poling period. The IR beams are independently mode-matched using lenses of focal length $f_{\mathrm{1a}}=-50$\,mm and $f_{\mathrm{1b}}=+50$\,mm in the $\lambda_1$ beam, and $f_{\mathrm{2a}}=-100$\,mm and $f_{\mathrm{2b}}=+100$\,mm in the $\lambda_2$ beam (as shown in Fig.~\ref{fig:fig1}(b)). The beams are subsequently overlapped on a dichroic optic $D_1$, and make a single pass of the PPSLT crystal. Optimum frequency conversion should occur when the confocal parameter $b$ of the input beams are related to the length of the nonlinear crystal $L$ by $\xi = L/b = 2.84$ \cite{Boyd1968}. However, the conversion efficiency is only weakly related to $\xi$, and focusing the beams too tightly (high $\xi$) puts a tighter constraint on the overlap of the beam waists in the axial direction, which is technically challenging to achieve. Instead we opt for a more relaxed $b$\,=\,32\,mm ($\xi$\,=\,1.56) for both input beams.  Note that $b$ is the confocal parameter of the beam inside the crystal, which is quite different to that of the beam in air due to the high refractive index of the material.  Our lens set-up allows for some adjustment of our chosen $\xi$ value, but we observe the conversion efficiency to be relatively insensitive to this parameter.  After the crystal, dichroic mirrors ($D_{2,3}$ in Fig.~\ref{fig:fig1}(b)) are used to separate the red and IR beams, and the unused IR power is dumped. 

\begin{figure}
	\includegraphics[width=\columnwidth]{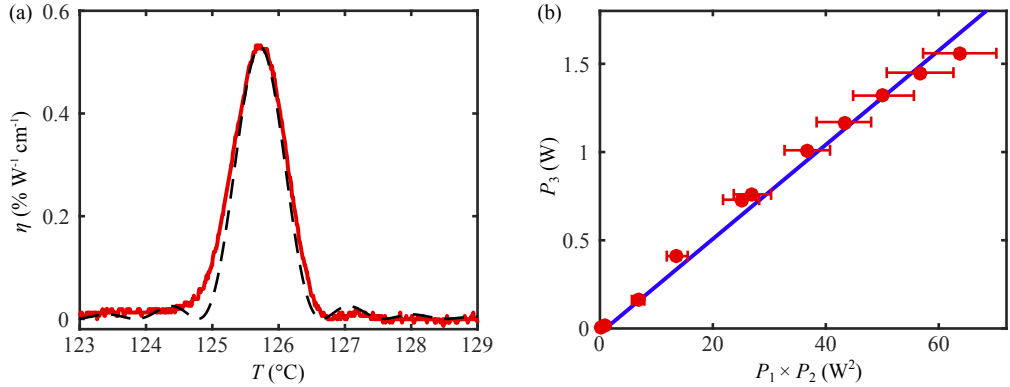}
	\caption{(a) Dependence of the SFG efficiency on PPSLT crystal temperature $T$.  Measured data are shown in red (a temperature independent background due to non-phase-matched SHG has been removed), with the prediction from Eq.~(\ref{eqn:temp1}) and Eq.~(\ref{eqn:temp2}) shown in black dashes. A small temperature offset (0.55\,$^{\circ}$C) has been added to the prediction plot to match the experimental data. (b) Output power ($P_3$) of the SFG process for a range of input powers ($P_1 \times P_2$).  Data points are shown as red circles, and the blue line shows the least-squares linear fit to the data. }
\label{fig:fig2}
\end{figure}

The SFG relative conversion efficiency as a function of temperature $\eta'\left(T\right)$ is predicted to follow  a sinc$^2$ function:
\begin{equation}
\eta'\left(T\right) \propto \frac{\sin^2\left(\Delta k\left(T\right)L/2\right)}{\left(\Delta k\left(T\right)L/2\right)^2}\mbox{.}
\label{eqn:temp2}
\end{equation}
The measured temperature dependence of the conversion efficiency is plotted in Fig.~\ref{fig:fig2}(a) along with the expected curve from Eq.~(\ref{eqn:temp1}) and Eq.~(\ref{eqn:temp2}). We observe a feature with FWHM\,$\approx$\,1\,$^{\circ}$C, so the  0.1\,$^{\circ}$C temperature resolution of our temperature controller is only just sufficient to stabilize the temperature to the top of the peak. 

The power dependence of the SFG process is shown in Fig.~\ref{fig:fig2}(b). The output power $P_3$ is broadly linear with the product of the input powers $P_1$ and $P_2$. We obtain a maximum power of $P_3$\,=\,1.6\,W of $\lambda_3$ light from the SFG process. A least-squares linear fit to the data in Fig.~\ref{fig:fig2}(b) yields the normalized conversion efficiency $\eta$, 
\begin{equation}
\eta = \frac{P_3}{P_1 P_2 L} = \left(0.53 \pm 0.02\right)\% \mbox{W}^{-1}\mbox{cm}^{-1}\mbox{,}
\label{eqn:efficiency}
\end{equation}
where $P_{1,2,3}$ are the beam powers of $\lambda_{1,2,3}$ and $L$ is the length of the PPSLT crystal.  This measured value is favourable in comparison to other reported CW conversion efficiencies for PPSLT, which are typically $\sim$\,0.3$\,\%$ to 0.4$\,\%$\,W$^{-1}$\,cm$^{-1}$ \cite{Hum2007,Sinha2008,Katz2004}.  However, it is noticeably worse than the normalized SFG conversion efficiencies achieved with the more widely used PPLN, which are typically $\sim$\,2.8$\,\%$\,W$^{-1}$\,cm$^{-1}$ \cite{Wilson2011,Hart1999}. The lower nonlinear coefficient $d_{33}$ of LT does not account for the total difference in conversion efficiency, which could be due to the greater difficulty of growing and poling stoichiometric lithium tantalate \cite{Hum2007}.  In the future, MgO doped stoichiometric LN could be a good compromise if the higher conversion efficiency is required, as the doping offers some level of protection against the damage mechanisms discussed above \cite{Kitamura2001,Furukawa2000}. 

\subsection{\label{sec:SHG}Second harmonic generation to the UV}
The majority of the light generated from the SFG process ($\lambda_3$) is coupled into a 10\,m long high power, polarization maintaining, single-mode optical fiber, with $>$\,65\,\% efficiency to a commercial frequency doubling unit (Toptica Photonics). A nonlinear crystal optimized for second harmonic generation using type-I phase matching is housed in a hermetically sealed bow-tie cavity. The unit contains a resonant electro-optic modulator (EOM) to apply sidebands to the $\lambda_3$ light for Pound-Drever-Hall (PDH) locking \cite{Drever1983} of the SHG cavity length by feedback to a piezo-mounted mirror. For a well-optimized system we have measured conversion efficiencies of $\sim$\,30\,\%, producing 230\,mW of UV light from 780\,mW of red light. During typical operation we obtain $>$\,100\,mW of UV power across a tuning range of 318.0\,nm to 319.3\,nm. The alignment of the SHG cavity requires re-optimization when the UV frequency is changed by more than $\sim$\,100\,GHz; this is a relatively quick and simple procedure. 

\subsection{\label{sec:laserlock}Laser frequency stabilization}
A tunable optical cavity is used to transfer the frequency stability of the 7.5\,kHz wide 5s$^2$\,$^1$S$_0$~$\rightarrow$~5s5p\,$^3$P$_1$ 689\,nm transition in $^{88}$Sr [Fig.~\ref{fig:fig1}(a)] to the UV laser system. An ECDL operating at 689\,nm is locked to a resonant mode of the optical cavity, and then the cavity length is stabilized to the 689\,nm sub-Doppler resonance feature in a thermal beam of Sr. The stable reference cavity can then be exploited to lock the frequency of the light at $\lambda_3$ [Fig.~\ref{fig:fig1}(b)]. 

Full details of the optical cavity design and construction are given in \cite{Boddy2014}. Briefly, the optical transfer cavity is 10.5\,cm long and consists of a ZERODUR\,\textregistered~spacer with a ring-shaped piezo between the spacer and each high reflectance mirror. One of the cavity piezos is a large-stroke piezo for long term drift correction, whist the other is a short-stroke piezo for faster length corrections \cite{Li2004}. The cavity is housed in a temperature stabilized vacuum chamber at $\sim$\,27\,$^{\circ}$C and $\sim$\,10$^{-7}$\,mbar. The finesse is $\left(41.3\,\pm\,0.6\right)$\,$\times$\,$10^3$ (FWHM\,=\,35\,kHz) at 689\,nm and $\left(109.9\,\pm\,0.4\right)$\,$\times$\,$10^3$ (FWHM\,=\,13\,kHz) at 638\,nm, as measured by cavity-ring-down spectroscopy. The 689\,nm laser is locked to the cavity using the PDH technique, and a portion of its output is used to generate an atomic error signal using sub-Doppler fluorescence spectroscopy in a Sr atomic beam. For this, the light is frequency modulated using a double-passed acousto-optic modulator and retro-reflected through the atomic beam. The resulting fluorescence is measured on a high gain photodiode, followed by demodulation using a lock-in amplifier to derive the error signal. Feedback is provided to both cavity piezos with different bandwidths, stabilizing the cavity length to the atomic reference signal such that the 689\,nm laser is locked on resonance. To lock the UV laser frequency, a small portion of the $\lambda_3$ light is picked off from the main beam path and overlapped with the 689\,nm beam on a dichroic filter ($D_4$ in Fig.~\ref{fig:fig1}(b)) providing alignment into the same optical cavity. The wavelength of the $\lambda_3$ light is locked to a resonant mode of the atom-referenced cavity using the PDH technique. A fast servo-loop controller (Toptica Photonics) provides feedback to both the diode current and piezo voltage of the $\lambda_1$ seed laser, with an overall loop bandwidth of $\sim$\,200\,kHz. The transfer cavity lock therefore both narrows the linewidth of the UV light as well as providing absolute frequency stability. Different modulation frequencies are used for the two PDH locks in order to prevent cross-talk between them. 

Since the atom-referenced cavity mode is not necessarily at a frequency that is useful for driving the Rydberg transitions, we use a fiber-coupled wideband EOM (JENOPTIK, PM635) to offset-lock the laser frequency from that of the cavity mode.  We use the ``electronic sideband'' technique \cite{Thorpe2008}, where a tunable sideband is locked to the cavity at a detuning $\nu_{\mathrm{c}}$ from the main laser frequency. The carrier frequency of the EOM ($\nu_{\mathrm{c}}$\,=\,0.1\,MHz~--~5\,GHz) is phase modulated at a frequency $\nu_{\mathrm{sb}}$\,=\,8\,MHz to produce the modulation sidebands required for PDH locking [Fig.~\ref{fig:fig1}(b)]. Demodulation at $\nu_{\mathrm{sb}}$ yields a standard PDH error signal, locking the laser at a frequency $\nu_{\mathrm{c}}$ away from the cavity mode. The UV laser frequency can be stepped or scanned while locked to the cavity by varying $\nu_{\mathrm{c}}$ \cite{Thorpe2008}. This allows us to manipulate the laser frequency with high precision during an experimental sequence, such as for the spectrum shown in Fig.~\ref{fig:fig4}(b). The laser remains locked (or relocks very quickly) for step sizes within the capture range of the PDH lock ($\sim$\,1\,MHz in our case). 

\subsection{Laser frequency characterization}
We characterize the long-term frequency instability of the 638\,nm and 689\,nm lasers using heterodyne beat measurements with a GPS-referenced optical frequency comb (Toptica Photonics). The fiber laser-based frequency comb has a difference frequency generation design that eliminates the carrier-envelope offset frequency. The tenth harmonic of the 80\,MHz repetition rate is locked to an ultra-low noise oven-controlled oscillator (OXCO), which in turn is locked to a 10\,MHz GPS-disciplined quartz oscillator (GPSDO) (Jackson Labs Fury) for long-term ($\gtrsim$\,0.1\,s) stability. Independent measurements carried out by Toptica Photonics indicate that the performance of the frequency comb is limited by the RF-reference for timescales larger than the inverse of the locking bandwidth ($\sim$~few kilohertz) \cite{TopticaPC}. Each CW laser is overlapped with the corresponding section of the frequency comb and the beat note between the laser and the comb is measured on a fast photodiode. Repeated measurements of the beat frequency ($f$) are made over a period of a few hours using a zero-deadtime counter (Tektronix FCA3003). From these measurements we calculate the overlapping Allan deviation \cite{Walls1999} of the fractional frequency error  $(f - \bar{f})/f_0$, where $\bar{f}$ is the mean beat frequency and $f_0$ is the laser frequency (e.g. $\sim$\,434.829\,THz for the 689\,nm laser). 

\begin{figure}
	\centering
	\includegraphics{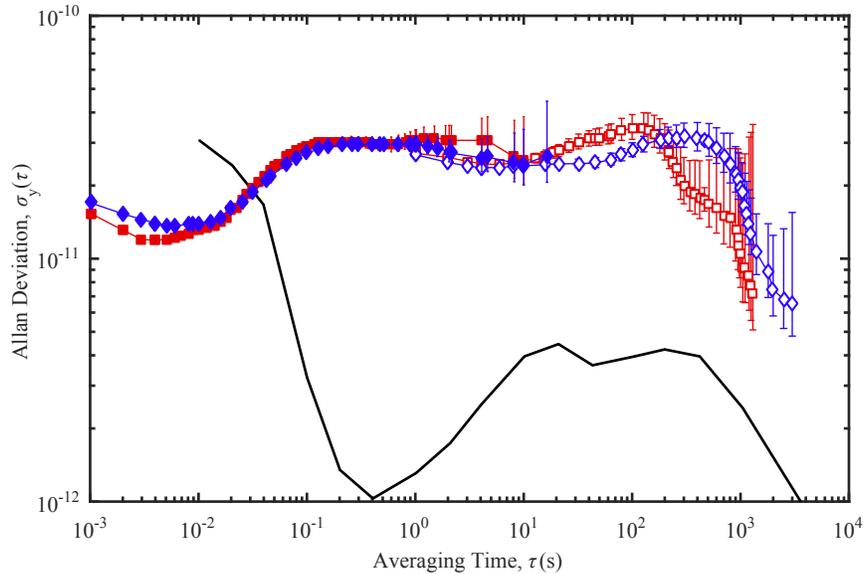}
	\caption{Allan deviation plots showing the fractional frequency instability of the 638\,nm (red squares) and 689\,nm (blue diamonds) lasers as measured by the optical frequency comb.  Measurements were made with either a 1\,ms or 1\,s frequency counter gate time, as indicated by filled and empty symbols respectively. The black line shows the specified instability of the GPSDO, reproduced with permission from Jackson Labs. }
\label{fig:fig3}
\end{figure}

The results are shown in Fig.~\ref{fig:fig3}, along with the manufacturer's data for the GPSDO. The Allan deviations for the 638\,nm and 689\,nm lasers are in reasonable agreement, which is expected because they are locked to the same optical reference cavity. A significant difference between these two measurements would indicate one or both of the lasers were not well locked to the transfer cavity. For $\tau\,\lesssim\,0.1\,$s, the beat frequency has less noise than the GPSDO, reflecting the much lower noise of the OXCO. Above $\tau\,\approx$\,0.1\,s the instability of the optical frequency comb should follow that of the GPSDO \cite{TopticaPC}. Instead, the measurements reveal excess frequency noise on both the CW lasers. We attribute this noise to the performance of the transfer cavity length servo to the atomic reference [Fig.~\ref{fig:fig1}(b)]. The low atomic fluorescence rate ($2\pi\,\times\,7.5$\,kHz) imposes a low bandwidth on the cavity length servo ($\sim$\,10\,Hz), which combined with a relatively high cavity drift rate due to piezo relaxation (up to $\sim$\,1\,MHz/s), leads to significant fluctuations in the lock. Nevertheless, Fig.~\ref{fig:fig3} shows that the fractional frequency instability of both lasers systems is $<4 \times 10^{-11}$ for all measurement times between $10^{-3}$\,s and $\gtrsim$\,$10^3$\,s, corresponding to frequency deviations of just $\sim$\,35\,kHz for the UV laser and $\sim$\,15\,kHz for the 689\,nm laser. 
 
For measurement times $\tau\lesssim\,10^{-3}\,$s, the frequency deviation measurements eventually become limited by the short-term jitter of the frequency comb, so this method cannot be used to infer the short-term linewidth. Instead, we estimate the short-term frequency noise of the lasers relative to the optical transfer cavity by looking at the spectrum of the in-loop PDH error signal. Note that this measurement gives a relative frequency instability, rather than the absolute instability provided by the optical frequency comb, since the resonance frequency of the cavity may fluctuate due to acoustic vibrations and electrical noise on the mirror piezos. However, these cavity fluctuations will be relatively small at the timescales of interest for these measurements ($<$\,1\,ms). The in-loop error signal of the 638\,nm laser is converted into frequency deviations from the cavity mode center by dividing through by the PDH discriminator gradient $\mathcal{D} = V_{\mathrm{pp}}/\delta \nu_{\mathrm{cav}}$, where $V_{\mathrm{pp}}$ is the peak to peak voltage of the unlocked error signal and $\delta \nu_{\mathrm{cav}}$ is the cavity mode FWHM. The overlapping Allan deviation yields a relative fractional frequency instability of $<4 \times 10^{-12}$ for all timescales between 2\,$\upmu$s and 1\,ms, which corresponds to frequency deviations of $<$\,4\,kHz in the UV. Similar analysis of the 689\,nm laser in-loop error signal is detailed in \cite{Boddy2014}, where we estimate the short-term linewidth to be $\sim$\,1.5\,kHz. 

Therefore if the short-term fluctuations of the transfer cavity length are small, then the in-loop error signal analysis of the locked lasers shows short-term laser frequency fluctuations to be less than the natural linewidths of the 689\,nm and Rydberg transitions ($\approx$\,10\,kHz). The long-term analysis with the GPS referenced frequency comb shows frequency deviations outside of the natural linewidths, and so in future we plan to improve the long-term performance by locking to an ultra-stable optical reference cavity of the type used in optical atomic clocks. Despite this, due to Doppler and power broadening effects, the measured frequency instability is still sufficient to perform high resolution spectroscopy of Rydberg states, as we show in the next section. 

\section{\label{sec:spec}Atomic spectroscopy}
To demonstrate the versatility of this laser system, including the high tunability and narrow linewidth, we performed spectroscopy of a range of triplet Rydberg states in $^{88}$Sr. We use a two-step excitation scheme at 689\,nm and 319\,nm, as shown in  Fig.~\ref{fig:fig1}(a). Rydberg spectroscopy was carried out in a laser cooled sample of $\sim$\,$10^6$ $^{88}$Sr atoms. A detailed description of the apparatus can be found in \cite{Boddy2014}. Briefly, atoms from a thermal beam of Sr are slowed, cooled and trapped in a ``blue'' MOT using the 5s$^2$\,$^1$S$_0$~$\rightarrow$~5s5p\,$^1$P$_1$ transition at 461\,nm, where they have a temperature of $\sim$\,5\,mK. They are subsequently loaded into a ``broadband red'' MOT using the 689\,nm intercombination line. In the broadband MOT, the 689\,nm light is frequency modulated to broaden its spectrum by $\sim$\,4\,MHz, which ensures efficient transfer from the relatively hot blue MOT. At the end of the broadband cooling stage the atoms typically have a temperature of $\sim\,10\,\upmu$K and a density of $\sim\,10^{11}$\,cm$^{-3}$.

\begin{figure}
	\includegraphics{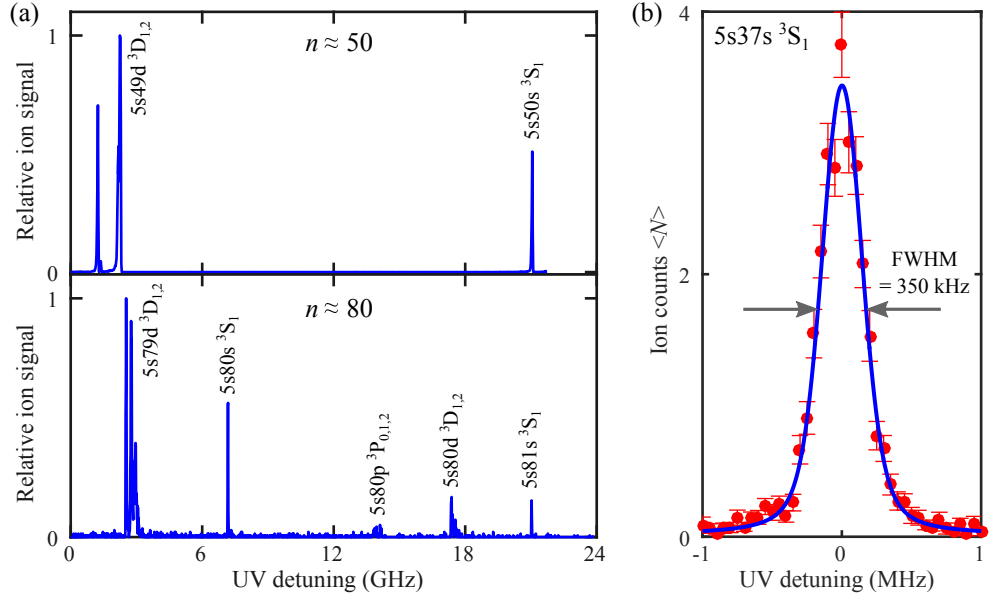}
	\caption{(a) Ion signal obtained from continuous scans of the UV detuning (measured relative to the start of the scan) in the regions around $n\,\approx\,50$ and $n\,\approx\,80$.  The intensities used for the excitation are $\sim$\,20\,mW\,cm$^{-2}$ and $\sim$\,1\,W\,cm$^{-2}$ for the 689\,nm and UV beams respectively, and each beam is larger than the size of the atom cloud. The variation in signal height across the scan is largely due to depletion of atoms from the MOT and is not an indication of transition strength. The frequency axis of the scans is calibrated on a 10\,MHz resolution wavemeter (High Finesse). The offset frequencies are are $\sim$\,940.649\,THz  ($n\,\approx\,50$) and $\sim$\,941.616\,THz ($n\,\approx\,80$). (b) High resolution scan of the $\mathrm{5s5p}\, ^\mathrm{3}\mathrm{P}_1,\,m_{J}=-1 \rightarrow \mathrm{5s37s}\, ^\mathrm{3}\mathrm{S}_1,\,m_{J}=0$ transition. The solid blue line shows the Voigt profile fit. The intensities used for the excitation are $\sim$\,0.1\,mW\,cm$^{-2}$ and $\sim$\,250\,mW\,cm$^{-2}$ for the 689\,nm and UV beams respectively, and each beam is larger than the size of the atom cloud. The central frequency of this feature is at $\sim$\,939.274\,THz. }
\label{fig:fig4}
\end{figure}

To locate individual Rydberg states, we perform continuous single-mode 24\,GHz scans of the UV laser frequency while the broadband MOT is operating. The UV laser is unlocked from the reference cavity, and scanned by piezo tuning of the $\lambda_1$ ECDL cavity length. The single-mode tuning range is ultimately limited by the $\sim40$ GHz bandwidth of the SFG process, although in practice the mode-hop free tuning range of the ECDL imposes a slightly lower limit. When the UV frequency matches a transition frequency, atoms are resonantly excited from the 5s5p\,$^3$P$_1$ state to a Rydberg state. Some of these Rydberg atoms spontaneously ionize, and we count the resulting ions using a multi-channel plate (MCP) and a fast oscilloscope. Results of such scans in the region $n\,\approx\,50$ and $n\,\approx\,80$ are shown in Fig.~\ref{fig:fig4}(a), where the Rydberg states appear as sharp peaks in the ion signal. We identify the Rydberg states by using the Rydberg-Ritz parameters given in \cite{Vaillant2012} to extrapolate from previous measurements at lower principal quantum numbers $n$ \cite{Beigang1982b,Rubbmark1978,Armstrong1979}. A small electric field ($\sim$\,1\,V\,cm$^{-1}$) applied during the excitation sequence, broadens and Stark splits the $^3$S$_1$- and $^3$D$_J$-states as well as allowing excitation to the $^3$P$_J$-states. Both effects are clearly visible in the $n\approx80$ data.

With the laser system locked to the transfer cavity (as described in section \ref{sec:laserlock}) we can perform Rydberg spectroscopy with $<$\,10\,kHz-level precision. Here, atoms are released from the broadband MOT, before being excited to the 5s37s\,$^3$S$_1$ Rydberg state using resonant 689~nm and 319~nm light. The 689~nm excitation beam counter-propagates at an angle of 30\,$^{\circ}$ to the UV beam, and both lasers are pulsed on simultaneously for 300\,$\upmu$s. A quantization magnetic field ($B$\,$\approx$\,0.14\,mT) applied during the excitation pulse separates the $m_J$ states. Following the Rydberg excitation, a 20\,$\upmu$s pulse of 408\,nm laser light is used to autoionize the Rydberg atoms and an electric field is applied to direct the ions towards the MCP \cite{Millen2010}. We repeat each measurement 10 times in the same atomic sample, after which the MOT is reloaded and the UV laser frequency stepped to the next detuning value by changing the EOM offset frequency $\nu_{\mathrm{c}}$. We take the spectrum in 50\,kHz steps, tuning the laser frequency up over the feature, then back down again. 

The results are shown in Fig.~\ref{fig:fig4}(b). The lineshape is very well described by a Voigt profile (reduced $\chi^2$\,=\,1.1) with an overall  FWHM of $\left(350\,\pm\,10\right)$\,kHz. From the fit we extract the FWHM of the inhomogeneous (Gaussian) $\left(280\,\pm\,30\right)$\,kHz and homogeneous (Lorentzian) $\left(120\,\pm\,30\right)$\,kHz contributions. Both are in agreement with our expected Doppler broadening and power broadening respectively. The fit yields the line center with a precision of $\sim$\,5\,kHz. Approximate transition frequencies are given for the observed states in the caption for Fig.~\ref{fig:fig4}. In the future, absolute frequency measurements at the $10^{-12}$ level will be possible with the use of the GPS-referenced optical frequency comb. However, this will require a detailed analysis of the electric and magnetic field induced line shifts, which is beyond the scope of this paper. 

\section{Conclusion}
In conclusion, we have developed a solid-state laser system for generating high power, narrow-linewidth UV light, which can be tuned across the range 316.3\,nm~--~317.7\,nm and 318.0\,nm~--~319.3\,nm. The Rydberg states shown in Fig.~\ref{fig:fig4} span a frequency range of over $2.3$\,THz, providing a clear illustration of the wide tunability of the laser design. At the same time, we demonstrate broad single-mode tuning range (24\,GHz) suitable for atomic spectroscopy, and precision control of the absolute frequency, enabling Rydberg spectroscopy with kilohertz precision. The long-term frequency instability of the laser system ($<$\,35\,kHz for timescales between $10^{-3}$\,s and $10^3$\,s) could be reduced by using an ultra-low drift optical reference cavity, and either an improved atomic spectroscopy set-up or the frequency comb as a long-term absolute reference. With these modifications, an absolute frequency instability below $1$\,kHz should be attainable. The laser output power is limited by the conversion efficiency of the PPSLT crystal. Although we obtain one of the highest conversion efficiencies so far reported for this material, replacement with MgO doped PPLN could provide a substantial increase in UV power. The precision and versatility of this laser system make it ideal for future Sr Rydberg experiments, including high precision spectroscopy of the Rydberg states and their interactions, and experiments involving Rydberg dressed potentials. The all solid-state design makes this an attractive replacement for complex frequency-doubled dye laser systems. 

\section*{Acknowledgments}
The authors would like to thank Thomas Puppe, Russell Kliese and Alexander Sell from Toptica Photonics for their work on developing and installing the optical frequency comb, and Chunyong Li at Durham University for maintaining it. We also thank Ian Hill from the National Physical Laboratory for comments on the manuscript. Financial support was provided by EPSRC grant EP/J007021/ and EU grants FP7-ICT-2013-612862-HAIRS and H2020-FETPROACT-2014-640378-RYSQ. The data presented in this paper are available for download \cite{Data}. 


\begin{thebibliography}{99}

\bibitem{Bollinger1985} J.~J.~Bollinger, J.~D.~Prestage, W.~M.~Itano, and D.~J.~Wineland, ``Laser-cooled-atomic frequency standard,'' Phys. Rev. Lett. \textbf{54}, 1000 (1985).
\bibitem{Hachisu2008} H.~Hachisu, K.~Miyagishi, S.~G.~Porsev, A.~Derevianko, V.~D.~Ovsiannikov, V.~G.~Pal’chikov, M.~Takamoto, and H.~Katori, ``Trapping of neutral mercury atoms and prospects for optical lattice clocks,'' Phys. Rev. Lett. \textbf{100}, 053001 (2008).
\bibitem{Lukin2001} M.~D.~Lukin, M.~Fleischhauer, R.~Cote, L. M.~Duan, D.~Jaksch, J.~I.~Cirac, and P.~Zoller, ``Dipole blockade and quantum information processing in mesoscopic atomic ensembles,'' Phys. Rev. Lett. \textbf{87}, 037901 (2001).
\bibitem{Pritchard2010} J.~D.~Pritchard, D.~Maxwell, A.~Gauguet, K.~J.~Weatherill, M.~P.~A.~Jones, and C.~S.~Adams, ``Cooperative atom-light interaction in a blockaded Rydberg ensemble,'' Phys. Rev. Lett. \textbf{105}, 193603 (2010).
\bibitem{Wilk2010} T.~Wilk, A.~Ga$\ddot{\mbox{e}}$tan, C.~Evellin, J.~Wolters, Y.~Miroshnychenko, P.~Grangier, and A.~Browaeys, ``Entanglement of two individual neutral atoms using Rydberg blockade,'' Phys. Rev. Lett. \textbf{104}, 010502 (2010).
\bibitem{Isenhower2010} L.~Isenhower, E.~Urban, X.~L.~Zhang, A.~T.~Gill, T.~Henage, T.~A.~Johnson, T.~G.~Walker, and M.~Saffman, ``Demonstration of a neutral atom controlled-NOT quantum gate,'' Phys. Rev. Lett. \textbf{104}, 010503 (2010).
\bibitem{Dudin2012} Y.~O.~Dudin, L.~Li, F.~Bariani, and A.~Kuzmich, ``Observation of coherent many-body Rabi oscillations,'' Nat. Phys. {\bf 8}, 790-794 (2012). 
\bibitem{Abel2011} R.~P.~Abel, C.~Carr, U.~Krohn, and C.~S.~Adams, ``Electrometry near a dielectric surface using Rydberg electromagnetically induced transparency,'' Phys. Rev. A {\bf 84}, 023408 (2011).
\bibitem{Sedlacek2012} J.~A.~Sedlacek, A.~Schwettmann, H.~K$\ddot{\mbox{u}}$bler, R.~L$\ddot{\mbox{o}}$w, T.~Pfau, and J.~P.~Shaffer, ``Microwave electrometry with Rydberg atoms in a vapour cell using bright atomic resonances,'' Nat. Phys. {\bf 8}, 819-824 (2012).
\bibitem{Lochead2013} G.~Lochead, D.~Boddy, D.~P.~Sadler, C.~S.~Adams, and M.~P.~A.~Jones, ``Number-resolved imaging of excited-state atoms using a scanning autoionization microscope,'' Phys. Rev. A {\bf 87}, 053409 (2013).
\bibitem{Saffman2002} M.~Saffman and T.~G.~Walker, ``Creating single-atom and single-photon sources from entangled atomic ensembles,'' Phys. Rev. A {\bf 66}, 065403 (2002). 
\bibitem{Greene2000} C.~H.~Greene, A.~S.~Dickinson, and H.~R.~Sadeghpour, ``Creation of polar and nonpolar ultra-long-range Rydberg molecules,'' Phys. Rev. Lett. \textbf{85}, 2458 (2000).
\bibitem{Boisseau2002} C.~Boisseau, I.~Simbotin, and R.~C$\hat{\mbox{o}}$t$\acute{\mbox{e}}$, ``Macrodimers: Ultralong range Rydberg molecules,'' Phys. Rev. Lett. {\bf 88}, 133004 (2002).
\bibitem{Bendkowsky2009} V.~Bendkowsky, B.~Butscher, J.~Nipper, J.~P.~Shaffer, R.~L$\ddot{\mbox{o}}$w, and T.~Pfau, ``Observation of ultralong-range Rydberg molecules,'' Nature {\bf 458}, 1005-1008 (2009). 
\bibitem{DeSalvo2015} B.~J.~DeSalvo, J.~A.~Aman, F.~B.~Dunning, T.~C.~Killian, H.~R.~Sadeghpour, S.~Yoshida, and J.~Burgd$\ddot{\mbox{o}}$rfer, ``Ultra-long-range Rydberg molecules in a divalent atomic system,'' Phys. Rev. A {\bf 92}, 031403(R) (2015).
\bibitem{Henkel2010} N.~Henkel, R.~Nath, and T.~Pohl, ``Three-dimensional roton excitations and supersolid formation in Rydberg-excited Bose-Einstein condensates,'' Phys. Rev. Lett. {\bf 104}, 195302 (2010).
\bibitem{Cinti2010} F.~Cinti, P.~Jain, M.~Boninsegni, A.~Micheli, P.~Zoller, and G.~Pupillo, ``Supersolid droplet crystal in a dipole-blockaded gas,'' Phys. Rev. Lett. {\bf 105}, 135301 (2010).
\bibitem{Gil2014} L.~I.~R.~Gil, R.~Mukherjee, E.~M.~Bridge, M.~P.~A.~Jones, and T.~Pohl, ``Spin squeezing in a Rydberg lattice clock,'' Phys. Rev. Lett. {\bf 112}, 103601 (2014).
\bibitem{Keating2015} T.~Keating, R.~L.~Cook, A.~M.~Hankin, Y.-Y.~Jau, G.~W.~Biedermann, and I.~H.~Deutsch, ``Robust quantum logic in neutral atoms via adiabatic Rydberg dressing,'' Phys. Rev. A \textbf{91}, 012337 (2015).
\bibitem{Mohapatra2007} A.~K.~Mohapatra, T.~R.~Jackson, and C.~S.~Adams, ``Coherent optical detection of highly excited Rydberg states using electromagnetically induced transparency,'' Phys. Rev. Lett. {\bf 98}, 113003 (2007).
\bibitem{Deiglmayr2006} J.~Deiglmayr, M.~Reetz-Lamour, T.~Amthor, S.~Westermann, A.~L.~de~Oliveira, and M.~Weidem$\ddot{\mbox{u}}$ller, ``Coherent excitation of Rydberg atoms in an ultracold gas,'' Opt. Commun. {\bf 264}, 293-298 (2006).
\bibitem{Thoumany2009} P.~Thoumany, T.~H$\ddot{\mbox{a}}$nsch, G.~Stania, L.~Urbonas, and Th.~Becker, ``Optical spectroscopy of rubidium Rydberg atoms with a 297\,nm frequency-doubled dye laser,'' Opt. Lett. {\bf 34}(11), 1621-1623 (2009).
\bibitem{Hankin2014} A.~M.~Hankin, Y.~-Y.~Jau, L.~P.~Parazzoli, C.~W.~Chou, D.~J.~Armstrong, A.~J.~Landahl, and G.~W.~Biedermann, ``Two-atom Rydberg blockade using direct 6S to $n$P excitation,'' Phys. Rev. A \textbf{89}, 033416 (2014). 
\bibitem{Vaillant2012} C.~L.~Vaillant, M.~P.~A.~Jones, and R.~M.~Potvliege, ``Long-range Rydberg-Rydberg interactions in calcium, strontium and ytterbium,'' J. Phys. B {\bf 45}(13), 135004 (2012).
\bibitem{Ovsyanikov2011} V.~D.~Ovsyanikov, A.~Derevianko, and K.~Gibble, ``Rydberg spectroscopy in an optical lattice: Blackbody thermometry for atomic clocks,'' Phys. Rev. Lett. \textbf{107}, 093003 (2011). 
\bibitem{DeSalvo2015b} B.~J.~DeSalvo, J.~A.~Aman, C.~Gaul, T.~Pohl, S.~Yoshida, J.~Burgd$\ddot{\mbox{o}}$rfer, K.~R.~A.~Hazzard, F.~B.~Dunning, and T.~C.~Killian, ``Rydberg-blockade effects in Autler-Townes spectra of ultracold strontium,'' arXiv:1510.08032.
\bibitem{Mukherjee2011}R.~Mukherjee, J.~Millen, R.~Nath, M.~P.~A.~Jones, and T.~Pohl, ``Many-body physics with alkaline-earth Rydberg lattices,'' J. Phys. B {\bf 44}(18), 184010 (2011).
\bibitem{Millen2010} J.~Millen, G.~Lochead, and M.~P.~A.~Jones, ``Two-electron excitation of an interacting cold Rydberg gas,'' Phys. Rev. Lett. {\bf 105}, 213004 (2010).
\bibitem{Millen2011} J.~Millen, G.~Lochead, G.~R.~Corbett, R.~M.~Potvliege, and M.~P.~A. Jones, ``Spectroscopy of a cold strontium Rydberg gas,'' J. Phys. B {\bf 44}(18), 184001 (2011).
\bibitem{Beigang1982b} R.~Beigang, K.~L$\ddot{\mbox{u}}$cke, D.~Schmidt, A.~Timmermann, and P.~J.~West, ``One-photon laser spectroscopy of Rydberg series from metastable levels in calcium and strontium,'' Phys. Scripta {\bf 26}, 183-188 (1982).
\bibitem{Manthey2014} T.~Manthey, T.~M.~Weber, T.~Niederpr$\ddot{\mbox{u}}$m, P.~Langer, V.~Guarrera, G.~Barontini, and H.~Ott, ``Scanning electron microscopy of Rydberg-excited Bose-Einstein condensates,'' New J. Phys. {\bf 16} (8), 083034 (2014).
\bibitem{Wilson2011} A.~C.~Wilson, C.~Ospelkaus, A.~P.~VanDevender, J.~A.~Mlynek, K.~R.~Brown, D.~Leibfried, and D.~J.~Wineland, ``A 750-mW, continuous-wave, solid-state laser source at 313\,nm for cooling and manipulating trapped $^9$Be$^+$ ions,'' Appl. Phys. B {\bf 105}(4), 741-748 (2011).
\bibitem{Rengelink2015} R.~J.~Rengelink, R.~P.~M.~J.~W.~Notermans, and W.~Vassen, ``A simple 2\,W continuous-wave laser system for trapping ultracold metastable helium atoms at the 319.8\,nm magic wavelength,'' arXiv:1511.00443.

\bibitem{Myers1997} L.~E.~Myers and W.~R.~Bosenberg, ``Periodically poled lithium niobate and quasi-phase-matched optical parametric oscillators,'' IEEE J. Quantum Elect. {\bf 33}(10), 1663-1672 (1997). 
\bibitem{Bruner2003} A.~Bruner, D.~Eger, M.~B.~Oron, P.~Blau, M.~Katz, and S.~Ruschin, ``Temperature-dependent Sellmeier equation for the refractive index of stoichiometric lithium tantalate,'' Opt. Lett. {\bf 28}(3), 194-196 (2003).
\bibitem{Hum2007} D.~S.~Hum, R.~K.~Route1, G.~D.~Miller, V.~Kondilenko, A.~Alexandrovski, J.~Huang, K.~Urbanek, R.~L.~Byer, and M.~M.~Fejer, ``Optical properties and ferroelectric engineering of vapor-transport-equilibrated, near-stoichiometric lithium tantalate for frequency conversion,'' J. Appl. Phys. {\bf 101}, 093108 (2007).
\bibitem{Kitamura2001} K.~Kitamura, Y.~Furukawa, S.~Takekawa, M.~Nakamura, A.~Alexandrovski, and M.~M.~Fejer, ``Optical damage and light-induced absorption in near-stoichiometric LiTaO$_3$ crystal,'' Lasers and Electro-Optics, 2001. CLEO~'01. Technical Digest. Summaries of papers presented at the Conference on; 02/2001, 138-139 (2001). 
\bibitem{Kim1969} Y.~S.~Kim and R.~T.~Smith, ``Thermal expansion of lithium tantalate and lithium niobate single crystals,'' J.~Appl. Phys. {\bf 40}, 4637 (1969).
\bibitem{Boyd1968} G.~D.~Boyd and D.~A.~Kleinman, ``Parametric interaction of focused Gaussian light beams,'' J. Appl. Phys. {\bf 39}, 3597, (1968).
\bibitem{Sinha2008} S.~Sinha, D.~S.~Hum, K.~E.~Urbanek, L.~Yin-wen M.~J.~F.~Digonnet, M.~M.~Fejer, and R.~L.~Byer, ``Room-temperature stable generation of 19 Watts of single-frequency 532-nm radiation in a periodically poled lithium tantalate crystal,'' J. Lightwave Technol. {\bf 26}(24), 3866-3871 (2008).
\bibitem{Katz2004} M.~Katz, R.~K.~Route, D.~S.~Hum, K.~R.~Parameswaran, G.~D.~Miller, and M.~M.~Fejer, ``Vapor-transport equilibrated near-stoichiometric lithium tantalate for frequency-conversion applications,'' Opt. Lett. {\bf 29}(15), 1775-1777 (2004). 
\bibitem{Hart1999} D.~L.~Hart, L.~Goldberg, and W.~K.~Burns, ``Red light generation by sum frequency mixing of Er/Yb fibre amplifier output in QPM LiNbO$_3$,'' Electron. Lett. {\bf 35}(1), 52-53 (1999).
\bibitem{Furukawa2000} Y.~Furukawa, K.~Kitamura, S.~Takekawa, A.~Miyamoto, M.~Terao, and N.~Suda, ``Photorefraction in LiNbO$_3$ as a function of [Li]/[Nb] and MgO concentrations,'' Appl. Phys. Lett. {\bf 77}, 2494 (2000). 

\bibitem{Drever1983} R.~W.~P.~Drever, J.~L.~Hall, F.~V.~Kowalski, J.~Hough, G.~M.~Ford, A.~J.~Munley, and H.~Ward, ``Laser phase and frequency stabilization using an optical resonator,'' Appl. Phys. B {\bf 31}(31), 97-105 (1983).
\bibitem{Boddy2014} D.~Boddy Doctoral Thesis, ``First observations of Rydberg blockade in a frozen gas of divalent atoms,'' Durham University \url{http://etheses.dur.ac.uk/10740/} (2014).
\bibitem{Li2004} Y.~Li, T.~Ido, T.~Eichler, and H.~Katori, ``Narrow-line diode laser system for laser cooling of strontium atoms on the intercombination transition,'' Appl. Phys. B {\bf 78}(3), 315-320 (2004).
\bibitem{Thorpe2008} J.~I.~Thorpe, K.~Numata, and J.~Livas, ``Laser frequency stabilization and control through offset sideband locking to optical cavities,'' Opt. Express {\bf 16}(20), 15980-15990 (2008).
\bibitem{TopticaPC} T.~Puppe, Toptica Photonics AG,  Lochhamer Schlag 19, 82166 Graefelfing (Munich), Germany (personal communication, 2015).
\bibitem{Walls1999} F.~L.~Walls and E.~S.~Ferre-Pikal, ``Frequency standards, characterization,'' Wiley Encyclopedia of Electrical and Electronics Engineering {\bf 12}, 767-775 (1999).
\bibitem{Rubbmark1978}  J.~R.~Rubbmark and S.~A.~Borgstr$\ddot{\mbox{o}}$m, ``Rydberg series in strontium found in absorption by selectively laser-excited atoms,'' Phys. Scr. {\bf 18}(4), 196-208 (1978).
\bibitem{Armstrong1979} J.~A.~Armstrong, J.~J.~Wynne, and P.~Esherick, ``Bound, odd-parity $J$~=~1 spectra of the alkaline earths: Ca, Sr, and Ba,'' J. Opt. Soc. Am. {\bf 69}(2), 211-230 (1979).

\bibitem{Data} E.~M.~Bridge, N.~C.~Keegan, A.~D.~Bounds, D.~Boddy, D.~P.~Sadler, and M.~P.~A.~Jones, ``A tunable CW UV laser with $<$35\,kHz absolute frequency instability for precision spectroscopy of Sr Rydberg states: supporting data,'' doi:10.15128/1v53jw96n. 

\end{thebibliography}
\end{document}